\documentclass[twocolumn,aps,prd,showpacs,preprintnumbers,floatfix,nofootinbib]{revtex4}

\usepackage{graphicx}

\newcommand{\be}{\begin{eqnarray}}
\newcommand{\ee}{\end{eqnarray}}

\begin{document}

%\preprint{gr-qc/yymmddd}

\title{A Complete Statistical Analysis for the Quadrupole Amplitude \\ in an Ellipsoidal Universe}

\author{A. Gruppuso}
\email{gruppuso@iasfbo.inaf.it}
\affiliation{INAF-IASF Bologna, 
Istituto di Astrofisica Spaziale e Fisica 
Cosmica di Bologna \\
Istituto Nazionale di Astrofisica, 
via Gobetti 101, I-40129 Bologna, Italy and \\
INFN, Sezione di Bologna, via Irnerio 46, I-40126 Bologna, Italy}

\date{\today} 

\begin{abstract} 
A model of Universe with a small eccentricity due to the presence of a magnetic field at the decoupling time (i.e. an Ellipsoidal Universe) has been recently proposed for the solution of the low quadrupole anomaly of the angular power spectrum of cosmic microwave background anisotropies.
%An Ellipsoidal Universe (i.e. a model of Universe with a small eccentricity due to the presence of a magnetic field at the decoupling time) has been recently proposed for the solution of the low quadrupole anomaly of the angular power spectrum of cosmic microwave background anisotropies.
We present a complete statistical analysis of that model showing that the probability of increasing of the amplitude of the quadrupole is larger than the probability of decreasing in the whole parameters' space. 
%As a consequence, with our unpriorized analysis, the low value for the observed quadrupole is more %unlikely in the considered treatment of Ellipsoidal Universe than in a standard $\Lambda$CDM model with %no eccentricity.

\end{abstract}


\maketitle

\section{Introduction}

The three year release of WMAP data \cite{Hinshaw} confirms that the amplitude of the Quadrupole of Angular Power Spectrum (APS) of Cosmic Microwave Background (CMB) anisotropies is approximately five times lower than the expected value of the $\Lambda$CDM model.

This anomaly 
\footnote{The Quadrupole anomaly is not the unique anomaly that is present at large angular scales of CMB maps. For other anomalies see for example \cite{Copi2006}, \cite{deOliveira-Costa2003}, \cite{Schwarz2004}, \cite{Copi2003}, \cite{Abramo2006}.} has attracted much interest and many papers have been published about this issue. The easiest explanation is that this low value could simply be a statistical fluke.
%(as the winning ticket in a lottery!). 
It is a matter of taste if this is enough satisfactory. Clearly this is not the case for many authors who investigated other possibilities as foreground or systematics not fully removed (see for example \cite{Abramo2003}, \cite{DSCamplitude}) or some effect of new physics (see \cite{Efstathiou2003}, \cite{Contaldi2003}, \cite{Piao2003}, \cite{Kawasaki2003}, \cite{Tsujikawa2003}, \cite{Moroi2003}, \cite{Gordon2004}, \cite{Weeks2003}, \cite{Piao2005}, \cite{Wu2006}).

Among these models, an Ellipsoidal Universe has been proposed as a model for the explanation of this anomaly \cite{campanelli}. 
It is very interesting to note that the presence of a magnetic field at decoupling time induces an eccentricity of the background metric which in turn modifies the energy of the CMB photons.
Following the treatment of \cite{campanelli}, this provides a novel temperature anisotropy that gives a contribution only to the quadrupole term (once expanded over Spherical Harmonics).
This effect has been considered and proposed in \cite{campanelli} to reconcile the observed quadrupole value with theoretical expectation of the $\Lambda$CDM model.
Other implications of this model can be found in \cite{paoloceapaper}.

The aim of the present paper is to associate a probability to the decreasing possibility given in this framework of Ellipsoidal Universe. 
We perform a complete statistical analysis of this model without constraining it to give the most favourite case (as done in literature).
We show that the observed quadrupole value is more unlikely in the considered treatment of Ellipsoidal Universe than in a standard $\Lambda$CDM model.

The paper is organized as follows: 
in Section \ref{ellipsoidal} we briefly describe the Ellipsoidal Universe model,
in Section \ref{statistics} technical details of the performed simulations are given and in Section \ref{conclusions} we draw our conclusions. 

\section{Ellipsoidal Universe}
\label{ellipsoidal}

It is shown in \cite{campanelli} that a small eccentricity $e_{\rm dec}$ at the 
decoupling time in the space-time metric provides a contribution only to the quadrupole terms ($\ell=2$) of CMB anisotropies
\begin{eqnarray}
\label{alme1} && a_{20}^{\rm e} = \frac{\sqrt{\pi}}{6\sqrt{5}} \,
             [1 + 3\cos(2 \vartheta) ] T_{cmb}\, e_{\rm dec}^2 \, , \\
\label{alme2} && a_{21}^{\rm e} = -(a_{2,-1}^{\rm e})^{*} =
             -\sqrt{\frac{\pi}{30}} \;
             e^{-i \varphi}  \sin(2\vartheta) T_{cmb}\, e_{\rm dec}^2 \, , \\
\label{alme3} && a_{22}^{\rm e} = (a_{2,-2}^{\rm e})^{*} =
             \sqrt{\frac{\pi}{30}} \; e^{-2 i \varphi} \sin^2\!\vartheta \,
             T_{cmb} \, e_{\rm dec}^2 \, ,
\end{eqnarray}
where ($\vartheta$,$\varphi$) represents the direction of the axis of the magnetic field that is responsible for the deviation from the perfect sphericity, $T_{cmb} \simeq 2.725$ K 
\cite{Mather} is the CMB temperature and the eccentricity $e_{\rm dec}$ is related to the magnetic field $B_0$ through the following equation
\be
e_{\rm dec} \simeq 10^{-2} h^{-1} {{\rm B_0} \over 10^{-8} {\rm G}}
\, ,
\ee
with ${\rm B_0}$ being the norm of the magnetic field at the present time and $h \simeq 72$ being the reduced (dimensionless) Hubble constant (implicitely defined by $H=\, h \, 100 \, {\rm km/s/Mpc}$).

These coefficients have to be added to the $a_{2 m}$ 
\footnote{We measure $a_{\ell m}$ in $\mu$K.} that are produced by the intrinsic, independent, Gaussian distributed temperature fluctuations of CMB in order to give the observed coefficients,
$a_{2 m}^{\rm obs}$:
\be
a_{2 m}^{\rm obs} = a_{2 m} + a_{2 m}^{\rm e}
\, .
\ee
Therefore the observed quadrupole 
\be
C^{{\rm obs}}_2 = \frac{1}{5} \sum_m a_{2 m}^{({\rm obs})} \left(a_{2 m}^{({\rm obs})}\right)^{\star}
\ee
is given by the following sum
\be
C^{{\rm obs}}_2 = C_2 + C_2^{mix} + C^{{\rm e}}_2 
\, ,
\label{observedquadrupole}
\ee
where $C_2$ is the intrinsic one, $C_2^{{\rm e}}$ is computed from Eqs.~(\ref{alme1}-\ref{alme3}) and is given
by
\be 
C_2^{{\rm e}} = 4 \pi \, T_{cmb}^2 \, e_{{\rm dec}}^4/225
\ee 
and $C_2^{mix}$ is the mixing term that is writable as
\be
C_2^{mix} = -2 f(\vartheta,\varphi) (C^{{\rm e}}_2)^{1/2}
\ee
with the function $f(\vartheta,\varphi)$ defined by
\begin{figure}
%\begin{tabular}{c}
%\includegraphics[scale=2]{parabola.eps}
\includegraphics{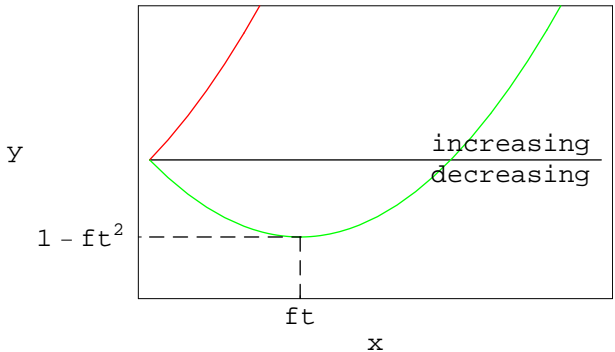}
%\end{tabular}
\caption{$y=y(x)$, see Eq.~(\ref{parabolic}). Branches for increasing or decreasing of the observed quadrupole. See also the text.}
\label{branches}
\end{figure}
\begin{widetext}
\be
f(\vartheta,\varphi) = \frac{-1}{4\sqrt{5}}{\left[ a_{20}\left(1+3\cos (2 \vartheta)\right)-2 {\sqrt{6}}\left( \left(a_{21}^{(R)}\cos \varphi - a_{21}^{(I)}\sin \varphi \right) \sin (2 \vartheta) - \left(a_{22}^{(R)}\cos (2 \varphi) - a_{22}^{(I)} \sin (2 \varphi)\right) \sin^2 \vartheta \right)\right]}
\, ,
\label{f}
\ee
\end{widetext}
where the labels $^{(R)}$ and $^{(I)}$ stand for the real and imaginary part 
of the intrinsic coefficients of the spherical harmonics respectively.
In this way Eq.~(\ref{observedquadrupole})
can be written as
\be
C^{\rm obs}_2 = C_2 -2f {C^{\rm e}_2}^{1/2} + C^{\rm e}_2
\, .
\label{observedquadrupole2}
\ee
Eq.~(\ref{observedquadrupole2}) tells us that the
observed quadrupole can be decreased with respect to the intrisic one
if the function $f$ is positive. 
This is easily seen if Eq.~(\ref{observedquadrupole2}) is rewritten 
as follows (in order to underline the parabolic behaviour)
\be
y = x^2 -2\,\tilde f \,x +1
\, ,
\label{parabolic}
\ee
where $y=C^{\rm obs}_2/C_2$, $x={C^{\rm e}_2}^{1/2}/C_2^{1/2}$ and $\tilde f = f /C_2^{1/2}$.
Eq.~(\ref{parabolic}) represents a parabolic behaviour with upward concavity.
Since $x>0$ it is clear that $y<1$ (i.e. a decreasing is obtainable for
the observed quadrupole) if and only if $\tilde f>0$ (that is the condition to have the abscissa 
of the vertex $x_V=\tilde f>0$). 
This is not always the case since $\tilde f$ can be positive or negative depending 
on the input values, as can be checked from Eq.~(\ref{f}).
In Fig.~\ref{branches} we plot Eq.~(\ref{parabolic}). 
The green branch represents Eq.~(\ref{parabolic}) for input values such that $\tilde f>0$ whereas
the red branch represents the parameter space for which $\tilde f<0$.
The horizontal black line divides the increasing from the decreasing $y$-region.
Both cases give a possible increasing of the quadrupole amplitude but only one case (the green branch) permits an interval of decreasing of the quadrupole amplitude.
As written in \cite{campanelli} the minimum is reached by 
$
x_{min}= \tilde f
$
%\, ,
%\label{xmin}
%\ee
that gives
%\be 
$y_{min}=1-\tilde f^2$.
%\, .
%\label{ymin}
%\ee
\section{Statistical Analysis}
\label{statistics}

It is possible to perform two kinds of analyses: the Minimum and the Full Shape Analysis. 
The first one, where the parameters are arbitrarely priorized such that the model is bounded
to $x_{min}=\tilde f$, is performed in \cite{campanelli}. This is done to maximize the effect in the 
direction we prefer. In the second case the analysis is faced in the full general case.
This is what is performed in the next subsection.

\subsection{Full Shape Analysis}
\label{fullshapeanalysis}
\begin{widetext}

\begin{figure}
%nomefile thetaphi.eps
% theta va 0 a 3
% phi va da 0 a 6
\includegraphics[width=4.3cm]{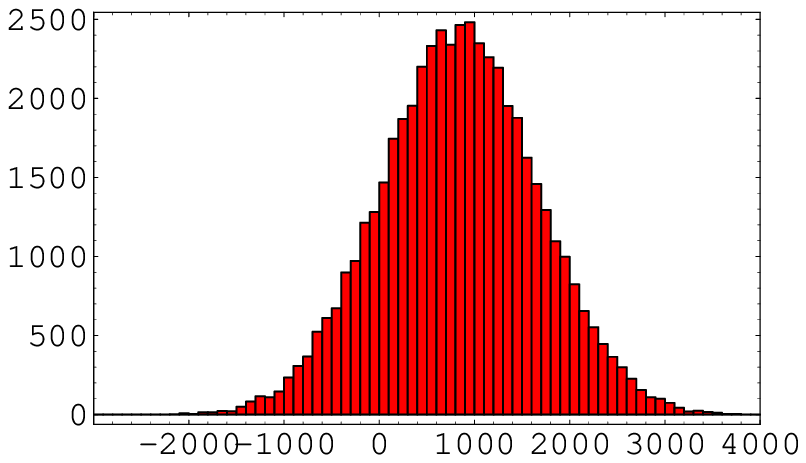}
\includegraphics[width=4.3cm]{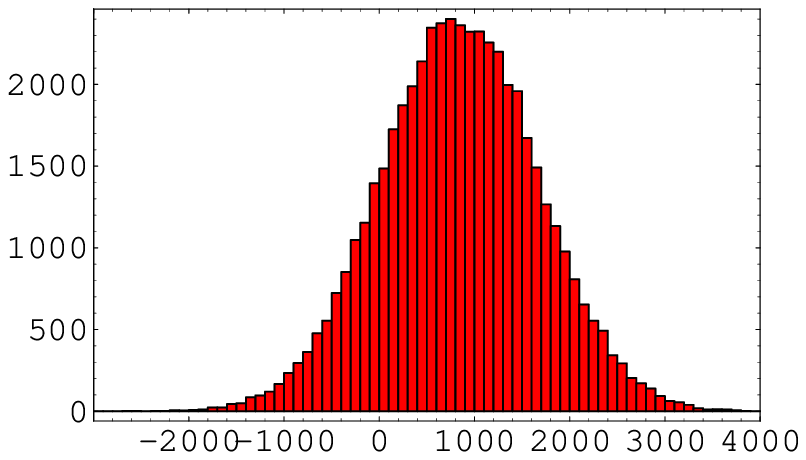}
\includegraphics[width=4.3cm]{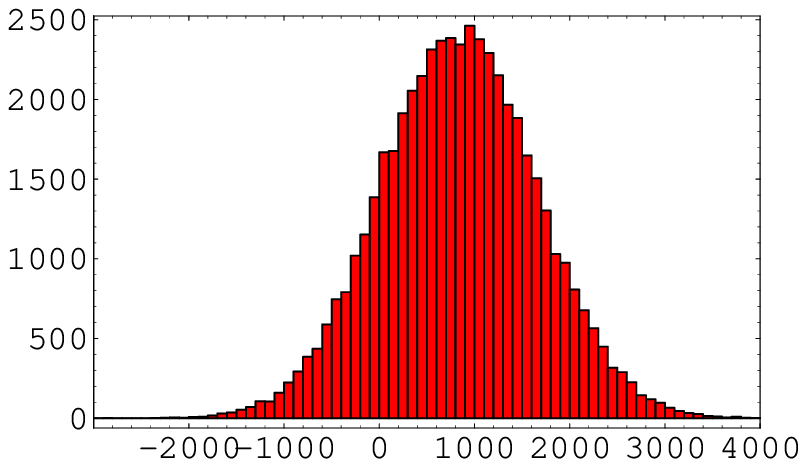}
\includegraphics[width=4.3cm]{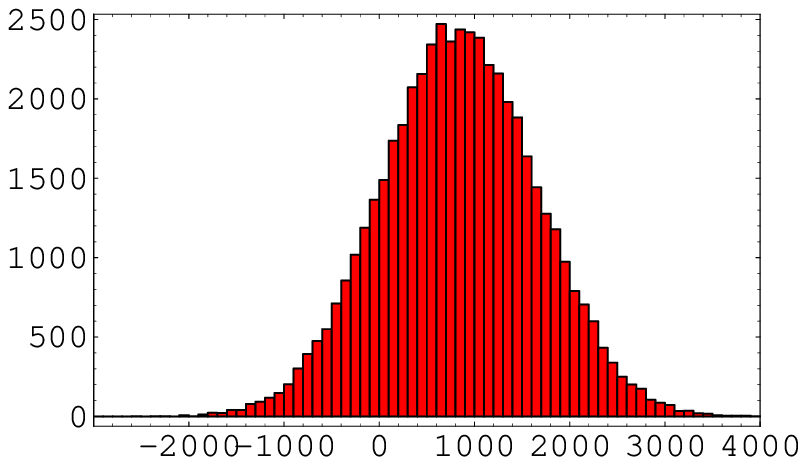}

\includegraphics[width=4.3cm]{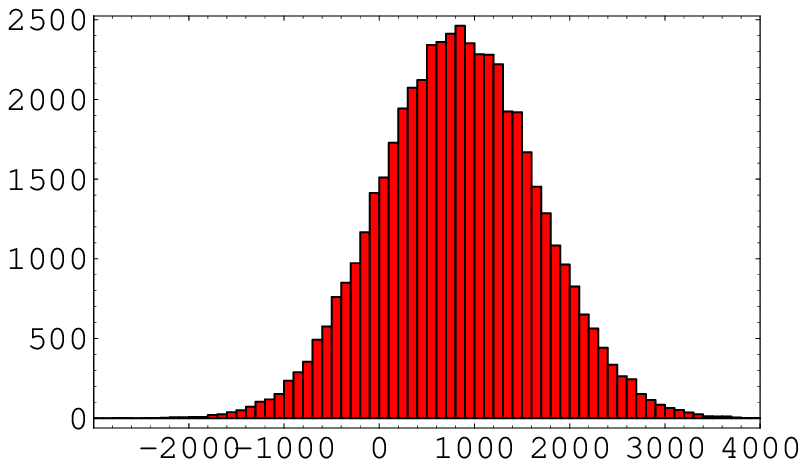}
\includegraphics[width=4.3cm]{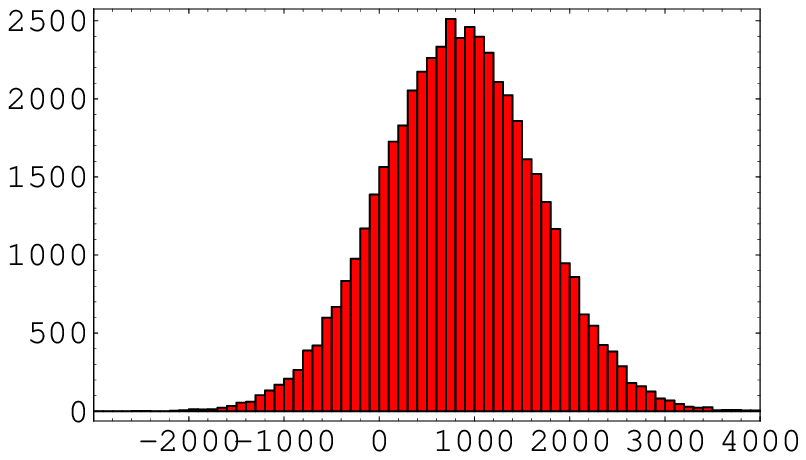}
\includegraphics[width=4.3cm]{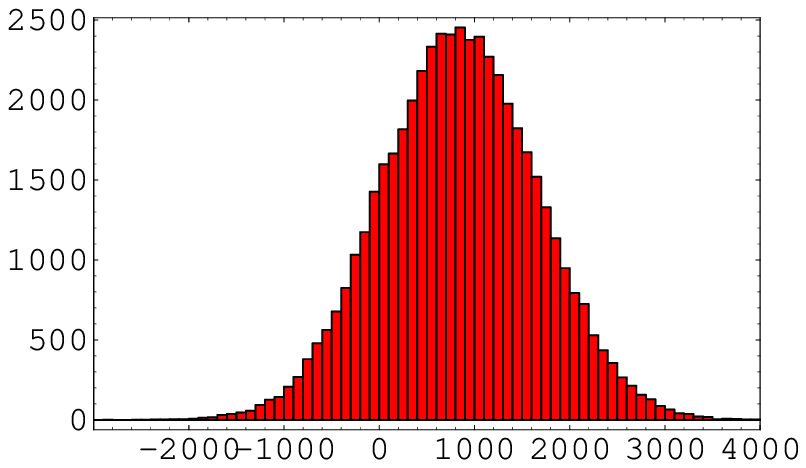}
\includegraphics[width=4.3cm]{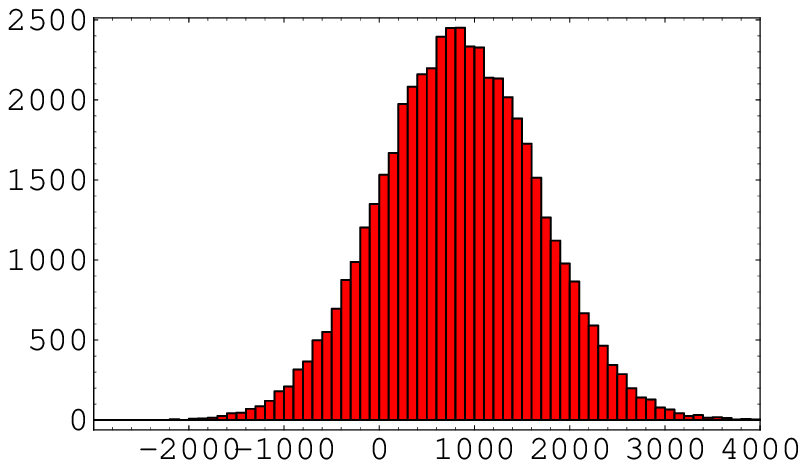}

\includegraphics[width=4.3cm]{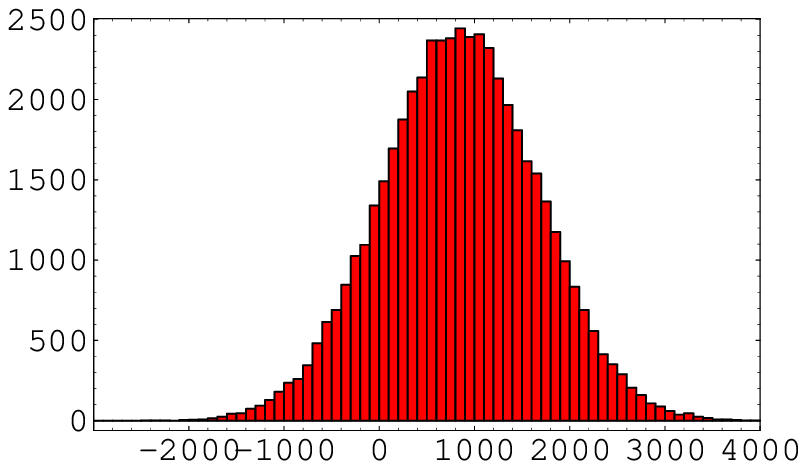}
\includegraphics[width=4.3cm]{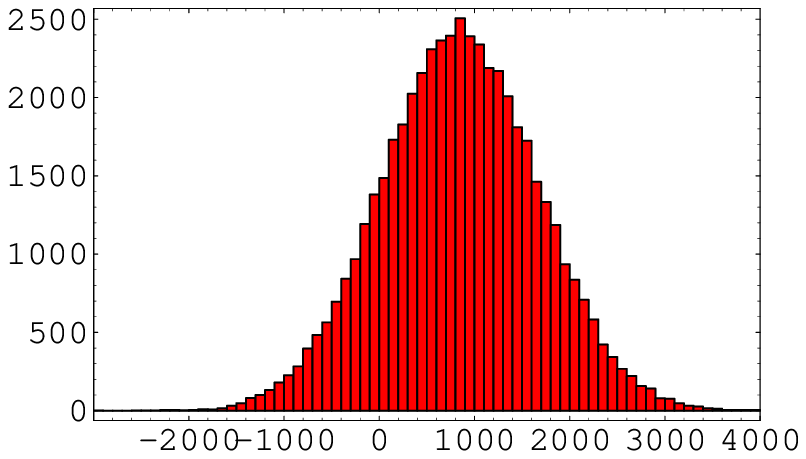}
\includegraphics[width=4.3cm]{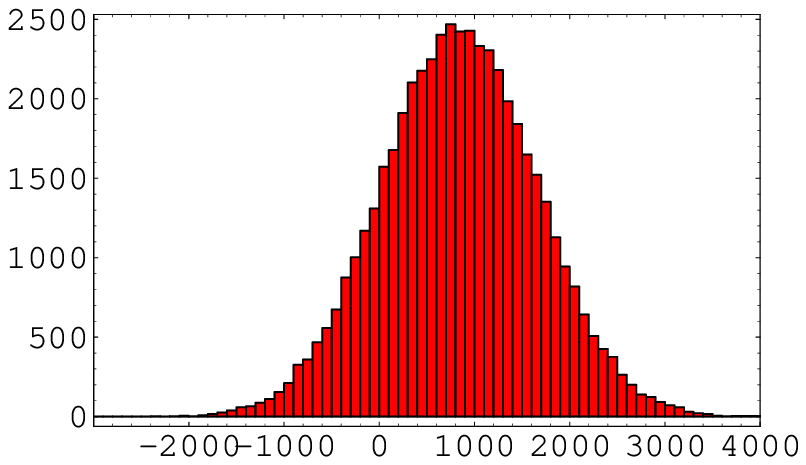}
\includegraphics[width=4.3cm]{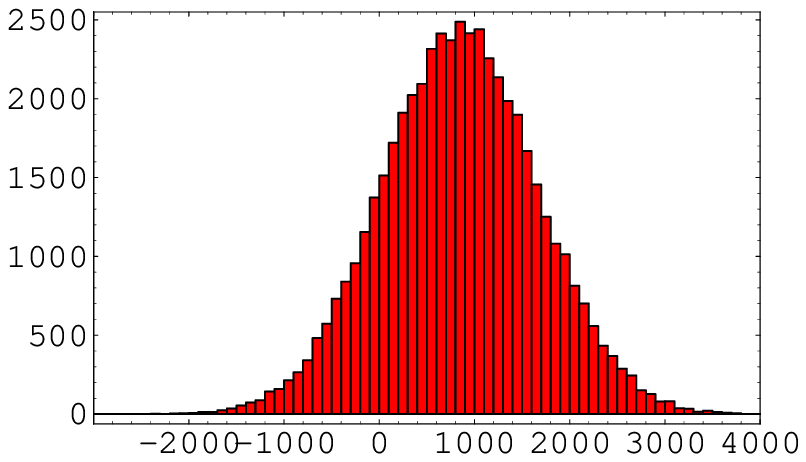}

\includegraphics[width=4.3cm]{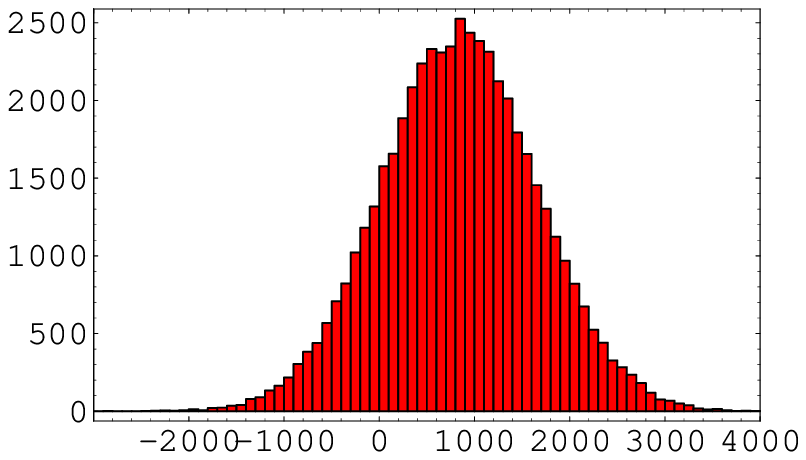}
\includegraphics[width=4.3cm]{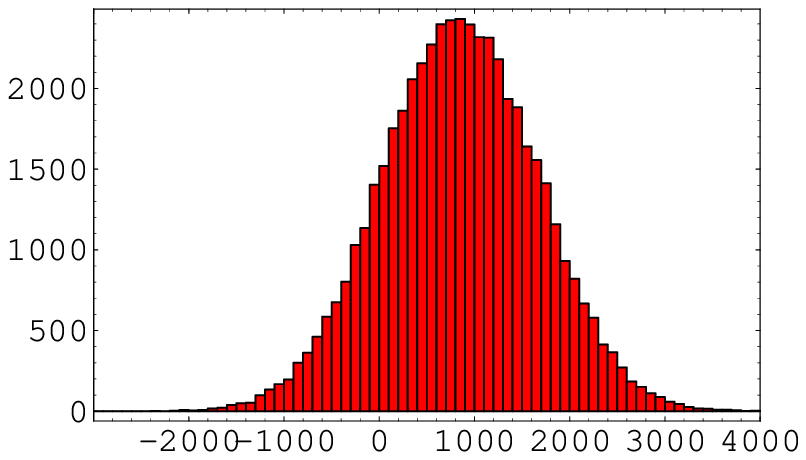}
\includegraphics[width=4.3cm]{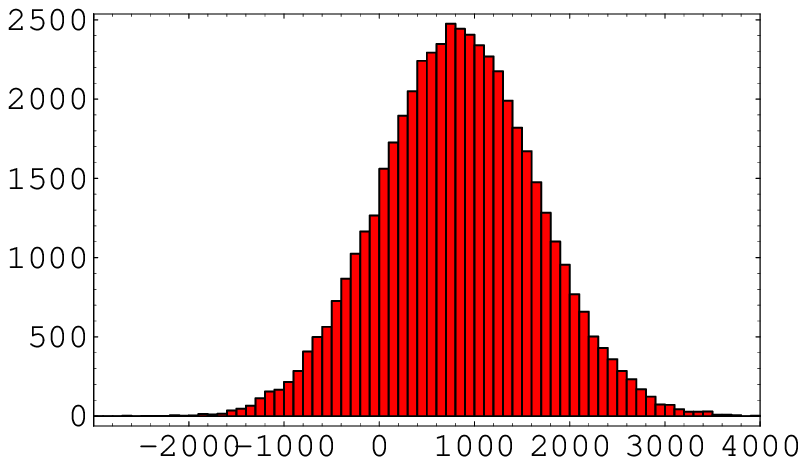}
\includegraphics[width=4.3cm]{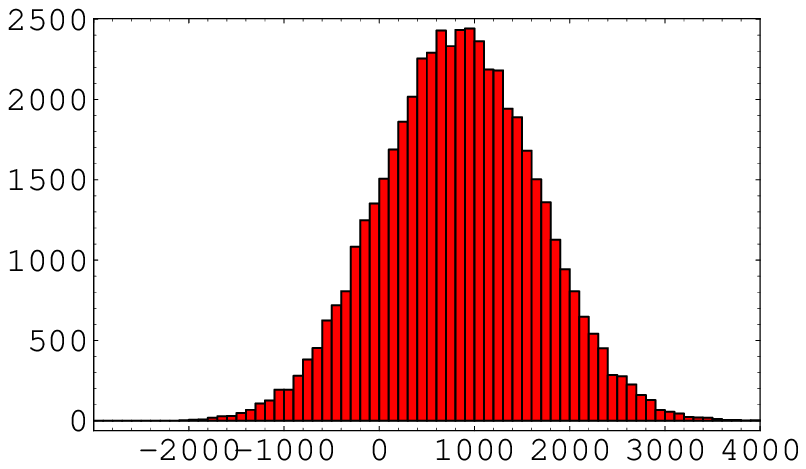}

\includegraphics[width=4.3cm]{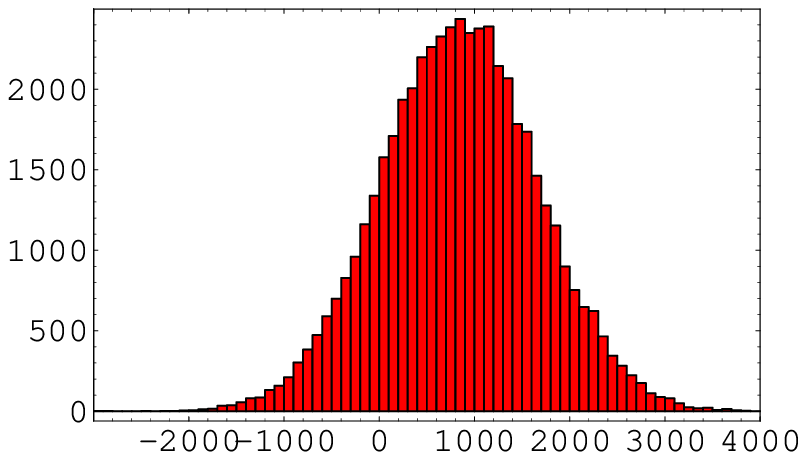}
\includegraphics[width=4.3cm]{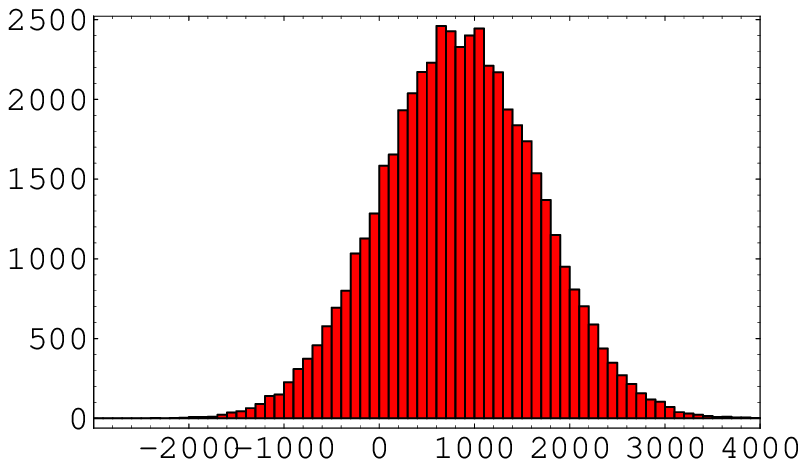}
\includegraphics[width=4.3cm]{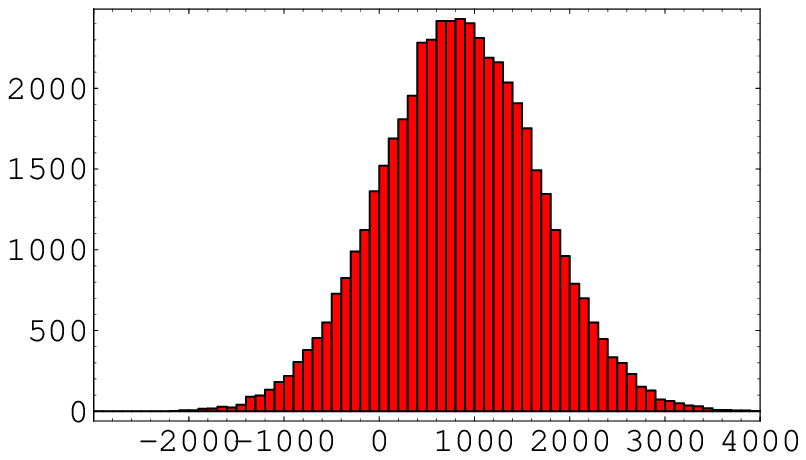}
\includegraphics[width=4.3cm]{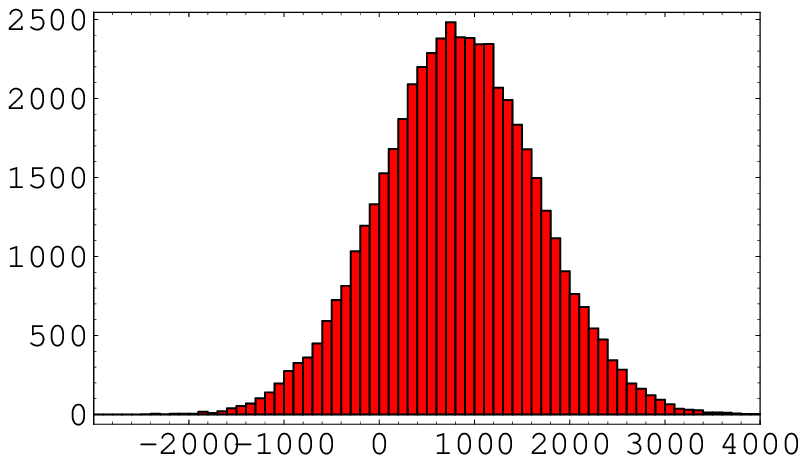}

\includegraphics[width=4.3cm]{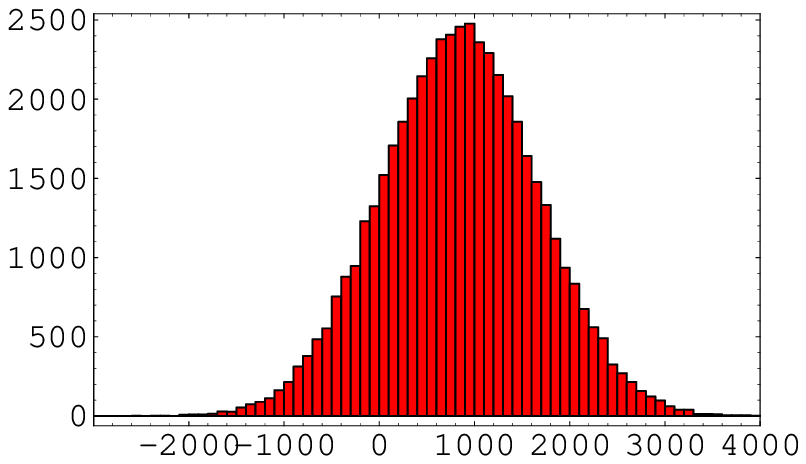}
\includegraphics[width=4.3cm]{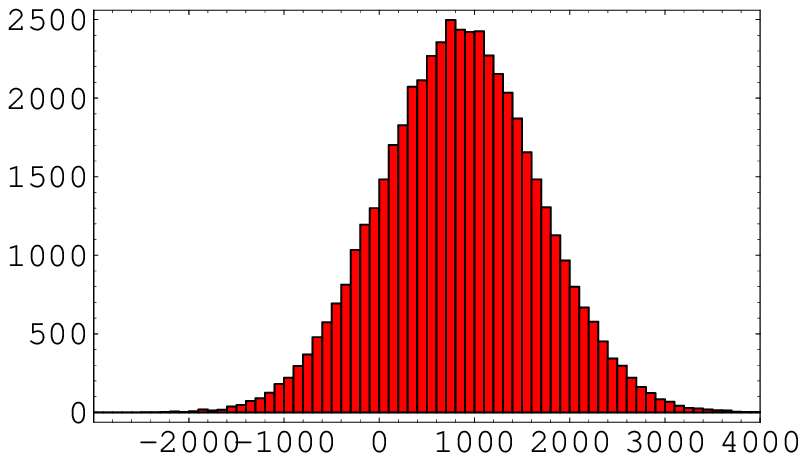}
\includegraphics[width=4.3cm]{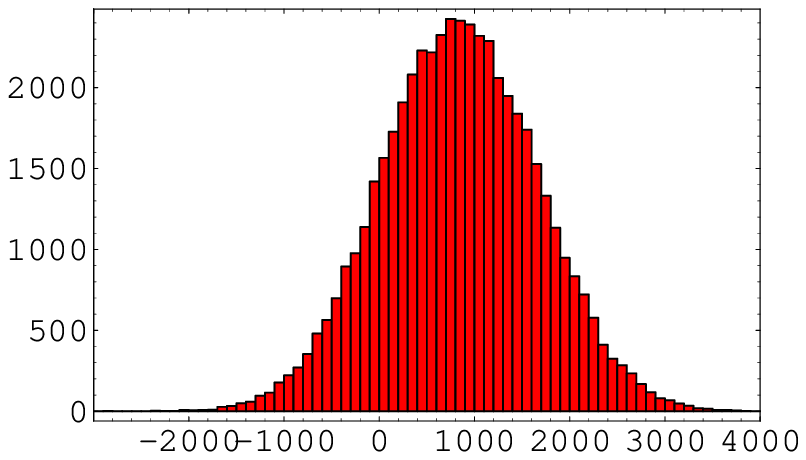}
\includegraphics[width=4.3cm]{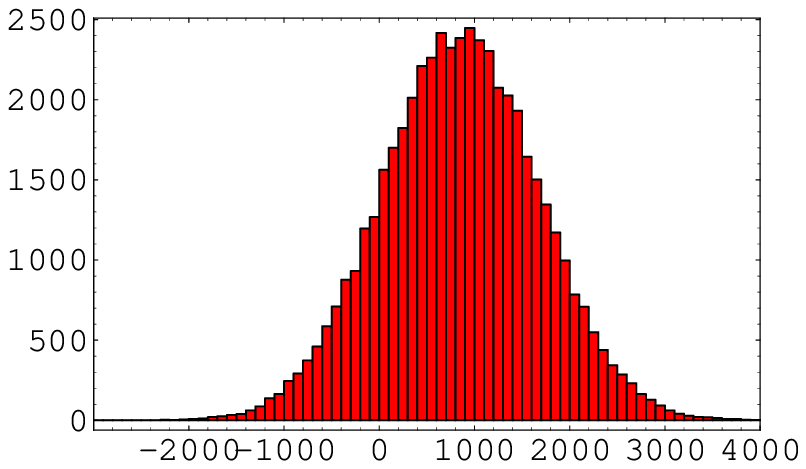}

\includegraphics[width=4.3cm]{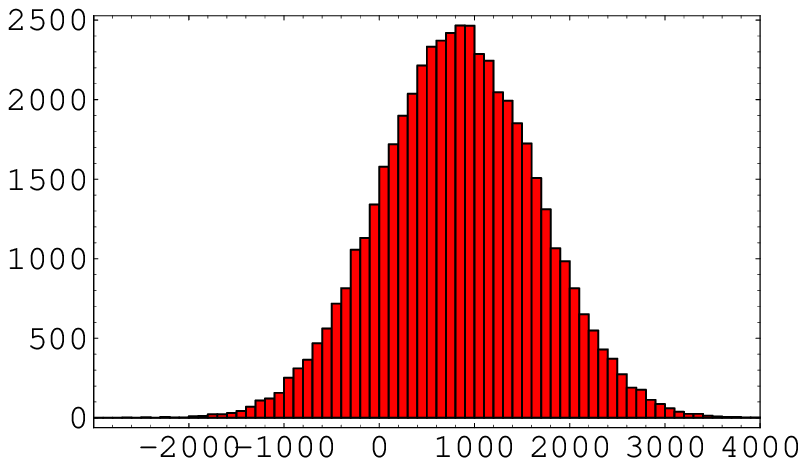}
\includegraphics[width=4.3cm]{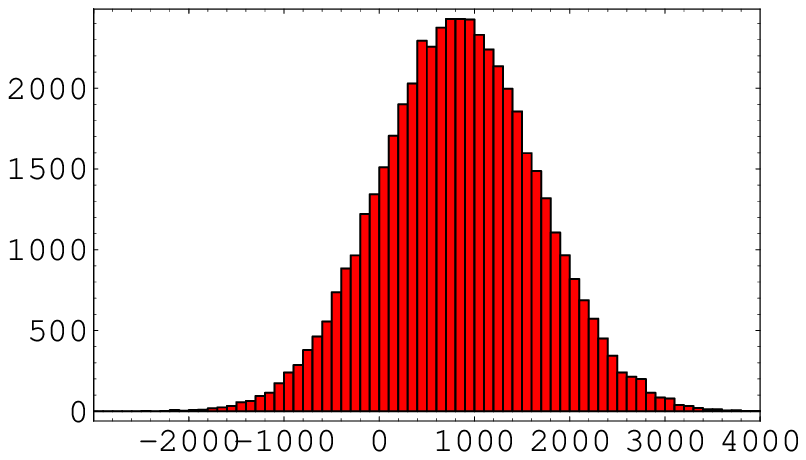}
\includegraphics[width=4.3cm]{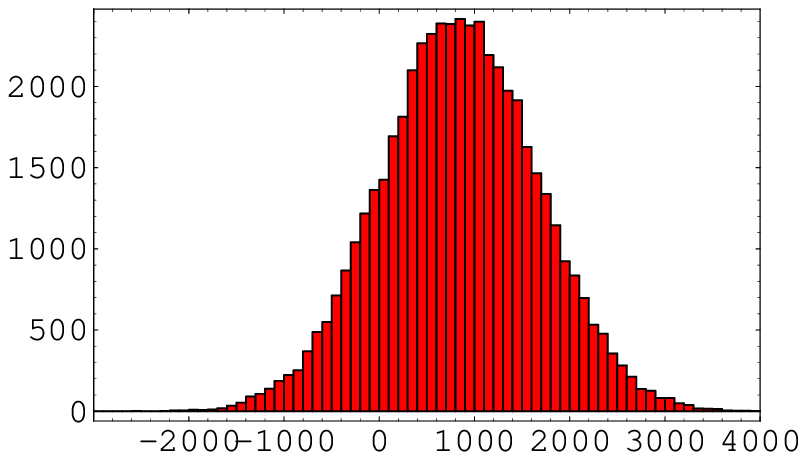}
\includegraphics[width=4.3cm]{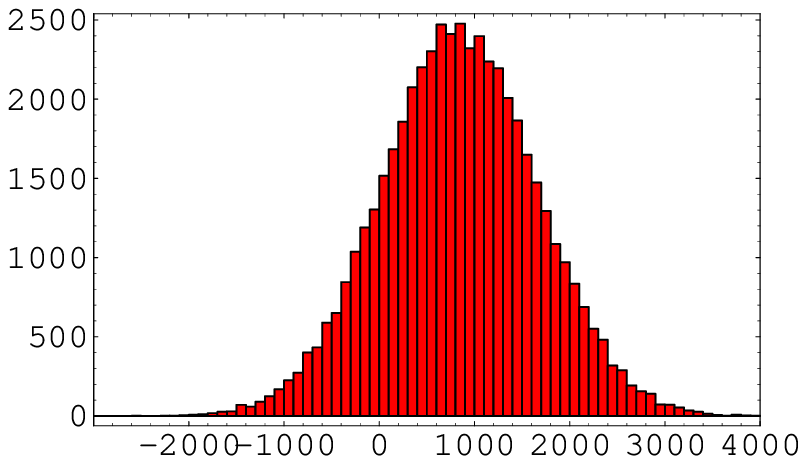}

%\end{tabular}
\caption{Likelihood (in terms of counts, y-axis) of $\delta C_2$ (x-axis, measured in $\mu$K$^2$). 
Panels in the same row have the same $\varphi=0$, $\pi/3$, $2\pi/3$, $\pi$, $4\pi/3$, $5\pi/3$ and 
$2\pi$ (in order from up to down). Panels in the same column have the same $\vartheta=0$, $\pi/3$, $2\pi/3$, $\pi$ (in order from left to right). The eccentricity is set to $e_{\rm dec}=0.67 \,\, 10^{-2}$.}
\label{probability}
\end{figure}

\end{widetext}

For each fixed eccentricity at decoupling $e_{\rm dec}$, and for each considered direction ($\vartheta$,$\varphi$), we perform $5 \times 10^3$ random extractions for the intrinsic quadrupole ($a_{2m}$) from a Gaussian distribution with zero mean and standard deviation $\sigma$
of the order of the expected quadrupole for the $\Lambda$CDM model, 
i.e. $\sigma \sim \sqrt{1000} \,\, \mu$K
\footnote{For the current purpose 
%it is not essential to take a more precise value 
%for $\sigma$, but 
it is sufficient an estimate of the order of magnitude for $\sigma$.}. These extractions are replaced in Eq.~(\ref{observedquadrupole2}) to obtain $C^{\rm obs}_2$ once the 
intrinsic quadrupole $C_2$ is computed.
This allows to obtain the likelihood of $\delta C_2 = C^{\rm obs}_2 - C_2$ for the fixed parameters 
$e_{\rm dec}$ and ($\vartheta$,$\varphi$). 
We consider the following values for $e_{\rm dec}=10^{-2}$, $0.5 \,\, 10^{-2}$ and $0.3 \,\, 10^{-2}$. 
Moreover we take into account $e_{\rm dec}=0.67 \,\, 10^{-2}$ that is the ``best case'' present in 
literature \cite{campanelli}. 
This is also the considered value for $e_{\rm dec}$ in all the panels of Fig.~\ref{probability} 
where we show the likelihood of $\delta C_2$.
The directional space of ($\vartheta$,$\varphi$) $\in [0,\pi]\times[0,2 \pi]$ is discretized with a 
step of $\pi/3$.
Fig.~\ref{probability} shows that the bell shape of the likelihood of $\delta C_2$ is always shifted 
towards positive values. 
This means that the increasing probability is always larger than the decreasing one. The same has 
been obtained for the other values of the eccentricity (that are not reported for sake of brevity).

\begin{widetext}

\begin{figure}
%nomefile thetaphi.eps
% theta va 0 a 3
% phi va da 0 a 6
\includegraphics[width=5.3cm]{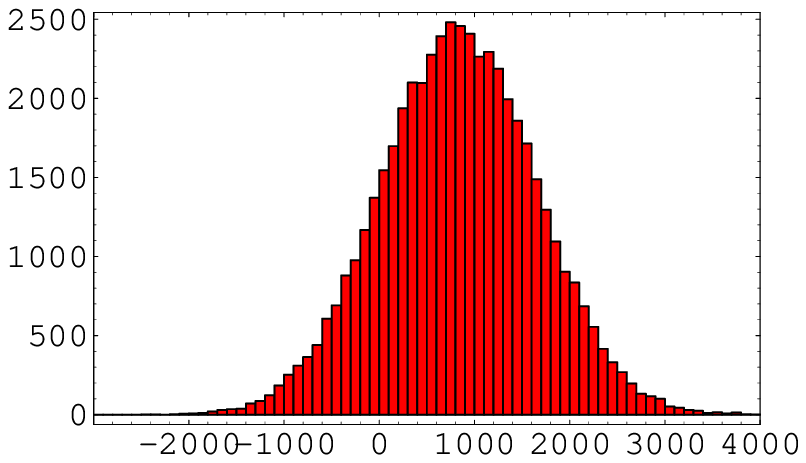}
\includegraphics[width=5.3cm]{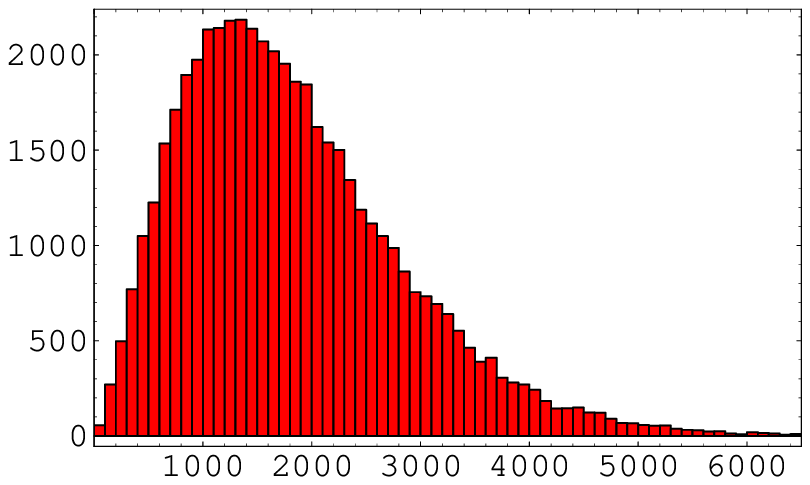}
\includegraphics[width=5.3cm]{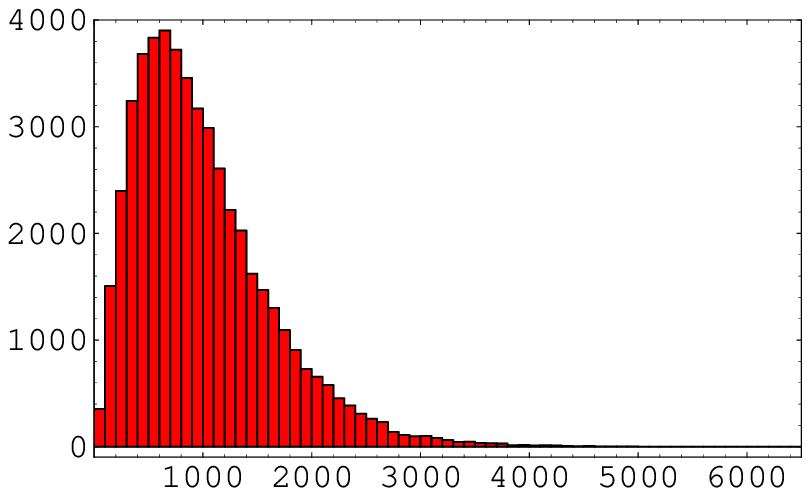}

\includegraphics[width=5.3cm]{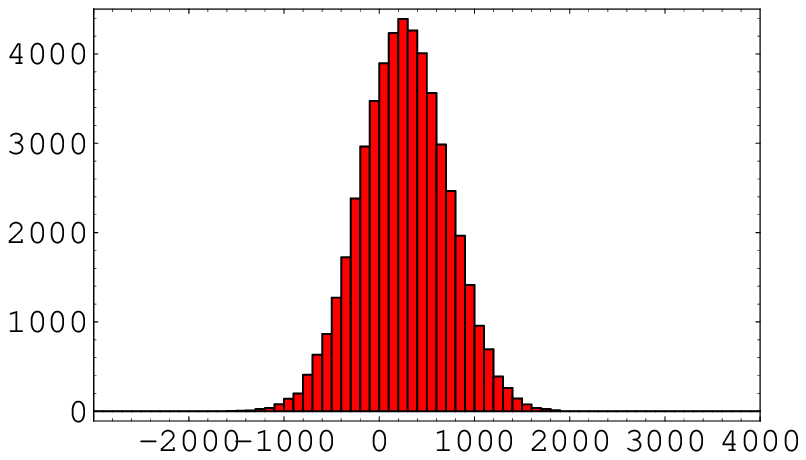}
\includegraphics[width=5.3cm]{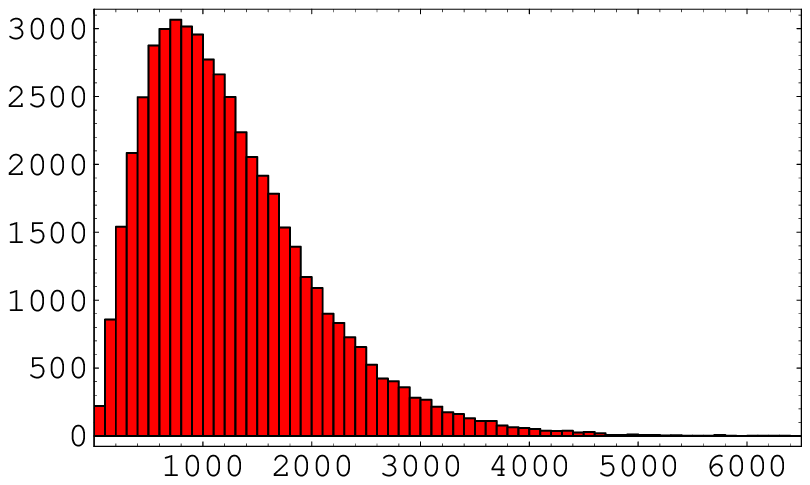}
\includegraphics[width=5.3cm]{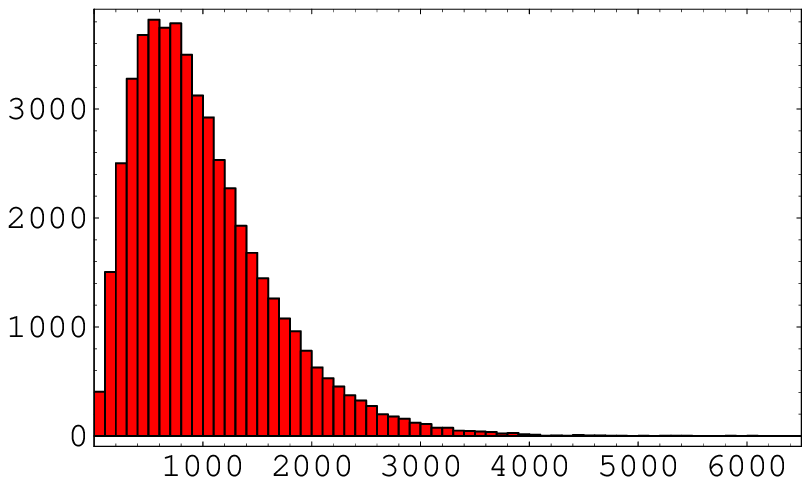}

\includegraphics[width=5.3cm]{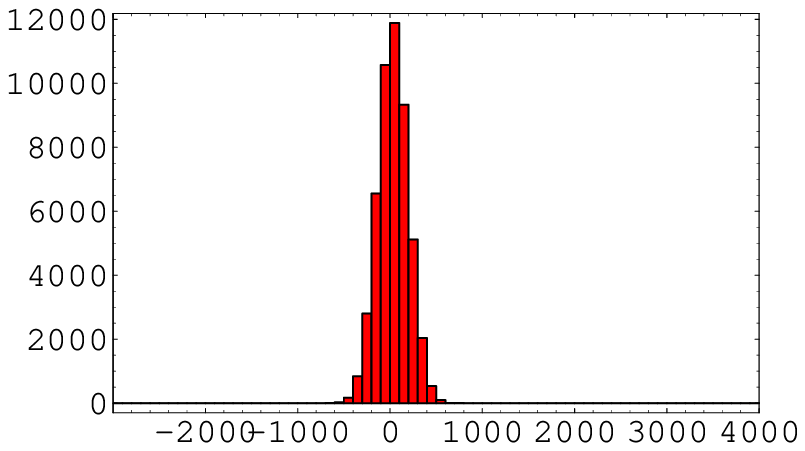}
\includegraphics[width=5.3cm]{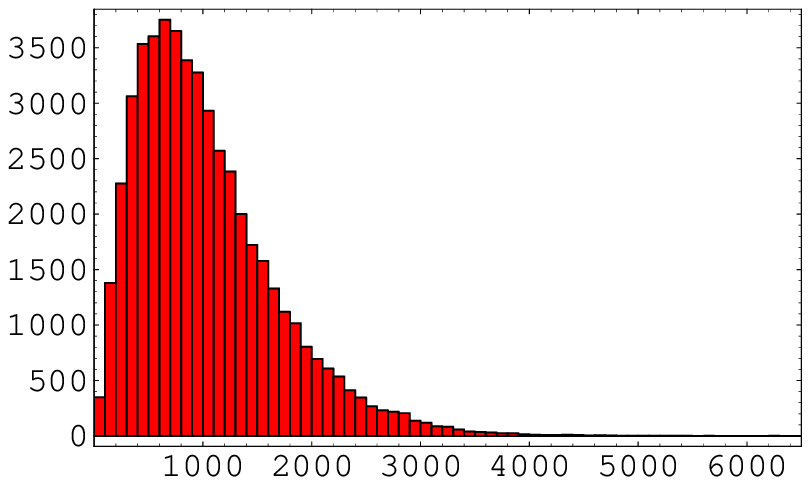}
\includegraphics[width=5.3cm]{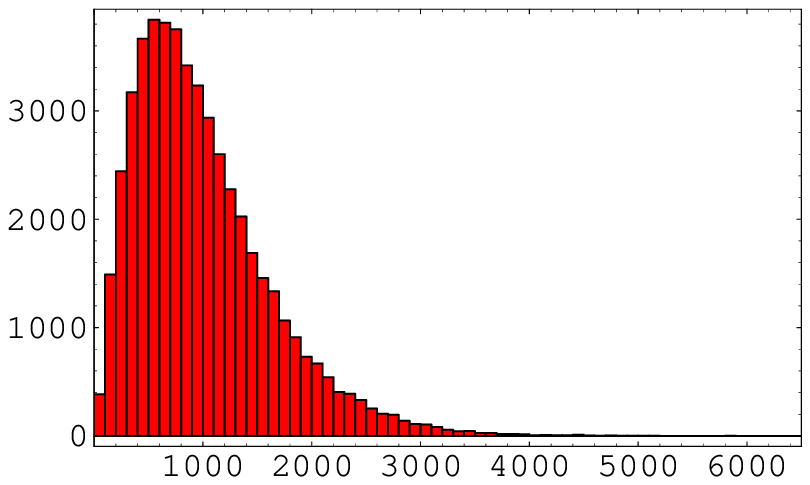}
%\includegraphics[width=5.3cm]{03C2observed.eps}
%\includegraphics[width=5.3cm]{03deltaC2.eps}
%\end{tabular}Likelihood (in terms of counts, y-axis) of $\delta C_2$ (x-axis, measured in $\mu$K$^2$)
\caption{Likelihood (in terms of counts, y-axis) of $\delta C_2$ (x-axis of the first column), $C^{\rm obs}_2$ (x-axis of the middle column) and $C_2$ (x-axis of the right column). $\delta C_2$, $C^{\rm obs}_2$ and $C_2$ are measured in $\mu$K$^2$. The first row refers to $e_{\rm dec}=0.67 \, 10^{-2}$,
the second row refers $e_{\rm dec}=0.5 \, 10^{-2}$ and the third refers $e_{\rm dec}=0.3 \, 10^{-2}$.
In all the panels $(\vartheta,\varphi)=(\pi/3,2 \pi/3)$.}
\label{probability2}
\end{figure}

\end{widetext}

In Fig.~\ref{probability2} we report the probability distribution for 
$\delta C_2$, $C^{\rm obs}_2$ and $C_2$ for $e_{\rm dec}=0.67 \,\, 10^{-2}$, $0.5 \,\, 10^{-2}$, 
$0.3 \,\, 10^{-2}$ at fixed $(\vartheta,\varphi)=(\pi/3,2\pi/3)$. 
Fig.~\ref{probability2} shows that the probability of extracting the observed WMAP value is smaller in the considered Ellipsoidal Universe than in a standard $\Lambda$CDM model with no eccentricity. Considering $C_2$(WMAP) $\sim 200 \mu$K$^2$ we compute
that for the observed quadrupole ($C^{\rm obs}_2$) the probability to obtain a smaller value is 
$0.7 \%$ (with $e_{\rm dec}=0.67 \,\, 10^{-2}$), $2.1 \%$ (with $e_{\rm dec}=0.5 \,\, 10^{-2}$) and 
$3.5 \%$ (with $e_{\rm dec}=0.3 \,\, 10^{-2}$) whereas for the intrinsic quadrupole ($C_2$) the probability is $3.8 \%$.
We end this section with the exptected value for the observed quadrupole (still for $(\vartheta,\varphi)=(\pi/3,2\pi/3)$ and $e_{\rm dec}=0.67 \,\, 10^{-2}$) that is computed to be 
$C^{\rm obs}_2 = 1822 \, \mu$K$^2$ whereas our intrinsic random extractions give 
$C_2 = 999 \, \mu$K$^2$.

%\begin{widetext}
%
%\begin{figure}
%nomefile thetaphi.eps
% theta va 0 a 3
% phi va da 0 a 6
%\includegraphics[width=5.3cm]{12deltaC2edec067.eps}
%\includegraphics[width=5.3cm]{12C2observededec067.eps}
%\includegraphics[width=5.3cm]{12C2intrinsicedec067.eps}
%
%\includegraphics[width=5.3cm]{12deltaC2edec05.eps}
%\includegraphics[width=5.3cm]{12C2observededec05.eps}
%\includegraphics[width=5.3cm]{12C2intrinsicedec05.eps}
%
%\includegraphics[width=5.3cm]{12deltaC2edec03.eps}
%\includegraphics[width=5.3cm]{12C2observededec03.eps}
%\includegraphics[width=5.3cm]{12C2intrinsicedec03.eps}
%\includegraphics[width=5.3cm]{03C2observed.eps}
%\includegraphics[width=5.3cm]{03deltaC2.eps}
%\end{tabular}Likelihood (in terms of counts, y-axis) of $\delta C_2$ (x-axis, measured in $\mu$K$^2$)
%\caption{Likelihood (in terms of counts, y-axis) of $\delta C_2$ (x-axis of the first column), $C^{\rm %obs}_2$ (x-axis of the middle column) and $C_2$ (x-axis of the right column). $\delta C_2$, $C^{\rm %obs}_2$ and $C_2$ are measured in $\mu$K$^2$. The first row refers to $e_{\rm dec}=0.67 \, 10^{-2}$,
%the second row refers $e_{\rm dec}=0.5 \, 10^{-2}$ and the third refers $e_{\rm dec}=0.3 \, 10^{-2}$.
%In all the panels $(\vartheta,\varphi)=(\pi/3,2 \pi/3)$.}
%\label{probability2}
%\end{figure}
%
%\end{widetext}

\section{Conclusions}
\label{conclusions}

We have statistically analyzed a model of Ellipsoidal Universe
recently proposed to solve the low quadrupole anomaly of CMB anisotropies.
%Considering the full shape of the observed quadrupole and not only the minimum,
Performing our unpriorized analisys, we find that the probability of increasing of 
the amplitude of the quadrupole is larger than the decreasing one.
We believe that this paper shows that the considered treatment of Ellipsoidal Universe cannot reconcile
current observations of the quadrupole amplitude of CMB anisotropies with theoretical predictions.
On the contrary in this model the observed quadrupole is more unlikely than in a
standard $\Lambda$CDM model with no eccentricity.

\newpage

{\bf Note added}
Before this paper was public, I have been informed of a different treatment of Ellipsoidal Universe \cite{paolocea} which should lead to a change in the sign of the function $f$ in Eq.~(\ref{f}).
This modification does not affect the results of this paper since for each point $(\vartheta,\varphi)$ of the considered parameters space, the number of extractions that give $f>0$ is close to the number of extractions that give $f<0$. 

{\bf Acknowledgments}
I thank C.~Burigana and F.~Finelli for many useful comments on the draft version of this paper. I wish to thank also L.~Campanelli, P.~Cea and L.~Tedesco for interesting correspondence.


\begin{thebibliography}{99}

\bibitem{Hinshaw}
  G.~Hinshaw {\it et al.}  [WMAP Collaboration],
  %``Three-year Wilkinson Microwave Anisotropy Probe (WMAP) observations:
  %Temperature analysis,''
  arXiv:astro-ph/0603451.
  %%CITATION = ASTRO-PH/0603451;%%

\bibitem{Copi2006}
  C.~Copi, D.~Huterer, D.~Schwarz and G.~Starkman,
  %``The Uncorrelated Universe: Statistical Anisotropy and the Vanishing Angular
  %Correlation Function in WMAP Years 1-3,''
  Phys.\ Rev.\  D {\bf 75}, 023507 (2007)
  [arXiv:astro-ph/0605135].
  %%CITATION = PHRVA,D75,023507;%%

\bibitem{deOliveira-Costa2003}
  A.~de Oliveira-Costa, M.~Tegmark, M.~Zaldarriaga and A.~Hamilton,
  %``The significance of the largest scale CMB fluctuations in WMAP,''
  Phys.\ Rev.\  D {\bf 69}, 063516 (2004)
  [arXiv:astro-ph/0307282].
  %%CITATION = PHRVA,D69,063516;%%

\bibitem{Schwarz2004}
  D.~J.~Schwarz, G.~D.~Starkman, D.~Huterer and C.~J.~Copi,
  %``Is the low-l microwave background cosmic?,''
  Phys.\ Rev.\ Lett.\  {\bf 93}, 221301 (2004)
  [arXiv:astro-ph/0403353].
  %%CITATION = PRLTA,93,221301;%%

\bibitem{Copi2003}
  C.~J.~Copi, D.~Huterer and G.~D.~Starkman,
  %``Multipole Vectors--a new representation of the CMB sky and evidence for
  %statistical anisotropy or non-Gaussianity at 2<=l<=8,''
  Phys.\ Rev.\  D {\bf 70}, 043515 (2004)
  [arXiv:astro-ph/0310511].
  %%CITATION = PHRVA,D70,043515;%%

\bibitem{Abramo2006}
  L.~R.~Abramo, A.~Bernui, I.~S.~Ferreira, T.~Villela and C.~A.~Wuensche,
  %``Alignment Tests for low CMB multipoles,''
  Phys.\ Rev.\  D {\bf 74}, 063506 (2006)
  [arXiv:astro-ph/0604346].
  %%CITATION = PHRVA,D74,063506;%%

\bibitem{Abramo2003}
  L.~R.~Abramo and L.~J.~Sodre,
  %``Can the Local Supercluster explain de low CMB multipoles?,''
  arXiv:astro-ph/0312124.
  %%CITATION = ASTRO-PH/0312124;%%

\bibitem{DSCamplitude}

%\bibitem{Burigana2006}
  C.~Burigana, A.~Gruppuso and F.~Finelli,
  %``On the dipole straylight contamination in spinning space missions dedicated
  %to CMB anisotropy,''
  Mon.\ Not.\ Roy.\ Astron.\ Soc.\  {\bf 371}, 1570 (2006)
  [arXiv:astro-ph/0607506].
  %%CITATION = MNRAA,371,1570;%%

%\bibitem{Gruppuso2006}
  A.~Gruppuso, C.~Burigana and F.~Finelli,
  %``Dipole Straylight Contamination and Low Multipoles,''
  PoS C {\bf MB2006}, 070 (2006)
  [arXiv:astro-ph/0607413].
  %%CITATION = POSCI,CMB2006,070;%%

\bibitem{Efstathiou2003}
  G.~Efstathiou,
  %``Is the low CMB quadrupole a signature of spatial curvature?,''
  Mon.\ Not.\ Roy.\ Astron.\ Soc.\  {\bf 343}, L95 (2003)
  [arXiv:astro-ph/0303127].
  %%CITATION = MNRAA,343,L95;%%

\bibitem{Contaldi2003}
  C.~R.~Contaldi, M.~Peloso, L.~Kofman and A.~Linde,
  %``Suppressing the lower Multipoles in the CMB Anisotropies,''
  JCAP {\bf 0307}, 002 (2003)
  [arXiv:astro-ph/0303636].
  %%CITATION = JCAPA,0307,002;%%

\bibitem{Piao2003}
  Y.~S.~Piao, B.~Feng and X.~m.~Zhang,
  %``Suppressing CMB quadrupole with a bounce from contracting phase to
  %inflation,''
  Phys.\ Rev.\  D {\bf 69}, 103520 (2004)
  [arXiv:hep-th/0310206].
  %%CITATION = PHRVA,D69,103520;%%

\bibitem{Kawasaki2003}
  M.~Kawasaki and F.~Takahashi,
  %``Inflation model with lower multipoles of the CMB suppressed,''
  Phys.\ Lett.\  B {\bf 570}, 151 (2003)
  [arXiv:hep-ph/0305319].
  %%CITATION = PHLTA,B570,151;%%

\bibitem{Tsujikawa2003}
  S.~Tsujikawa, R.~Maartens and R.~Brandenberger,
  %``Non-commutative inflation and the CMB,''
  Phys.\ Lett.\  B {\bf 574}, 141 (2003)
  [arXiv:astro-ph/0308169].
  %%CITATION = PHLTA,B574,141;%%

\bibitem{Moroi2003}
  T.~Moroi and T.~Takahashi,
  %``Correlated isocurvature fluctuation in quintessence and suppressed CMB
  %anisotropies at low multipoles,''
  Phys.\ Rev.\ Lett.\  {\bf 92}, 091301 (2004)
  [arXiv:astro-ph/0308208].
  %%CITATION = PRLTA,92,091301;%%

\bibitem{Gordon2004}
  C.~Gordon and W.~Hu,
  %``A Low CMB Quadrupole from Dark Energy Isocurvature Perturbations,''
  Phys.\ Rev.\  D {\bf 70}, 083003 (2004)
  [arXiv:astro-ph/0406496].
  %%CITATION = PHRVA,D70,083003;%%

\bibitem{Weeks2003}
  J.~Weeks, J.~P.~Luminet, A.~Riazuelo and R.~Lehoucq,
  %``Well-proportioned universes suppress CMB quadrupole,''
  Mon.\ Not.\ Roy.\ Astron.\ Soc.\  {\bf 352}, 258 (2004)
  [arXiv:astro-ph/0312312].
  %%CITATION = MNRAA,352,258;%%

\bibitem{Piao2005}
  Y.~S.~Piao,
  %``A Possible Explanation to Low CMB Quadrupole,''
  Phys.\ Rev.\  D {\bf 71}, 087301 (2005)
  [arXiv:astro-ph/0502343].
  %%CITATION = PHRVA,D71,087301;%%

\bibitem{Wu2006}
  C.~H.~Wu, K.~W.~Ng, W.~Lee, D.~S.~Lee and Y.~Y.~Charng,
  %``Quantum noise and a low cosmic microwave background quadrupole,''
  JCAP {\bf 0702}, 006 (2007)
  [arXiv:astro-ph/0604292].
  %%CITATION = JCAPA,0702,006;%%

\bibitem{campanelli}
  L.~Campanelli, P.~Cea and L.~Tedesco,
  %``Ellipsoidal Universe Can Solve The CMB Quadrupole Problem,''
  Phys.\ Rev.\ Lett.\  {\bf 97}, 131302 (2006)
  [Erratum-ibid.\  {\bf 97}, 209903 (2006)]
  [arXiv:astro-ph/0606266].
  %%CITATION = PRLTA,97,131302;%%
%\cite{Mather:1998gm}

\bibitem{paoloceapaper}
  P.~Cea,
  %``Ellipsoidal Universe Induces Large Scale CMB Polarization,''
  arXiv:astro-ph/0702293.
  %%CITATION = ASTRO-PH/0702293;%%

\bibitem{Mather}
  J.~C.~Mather, D.~J.~Fixsen, R.~A.~Shafer, C.~Mosier and D.~T.~Wilkinson,
  %``Calibrator Design for the COBE Far Infrared Absolute Spectrophotometer
  %(FIRAS),''
  Astrophys.\ J.\  {\bf 512} (1999) 511
  [arXiv:astro-ph/9810373].
  %%CITATION = ASJOA,512,511;%%

\bibitem{paolocea} 
P.~Cea, private communication.

\end{thebibliography}
\end{document}